\documentclass[reprint, showpacs, amsmath, amssymb, aps]{revtex4-1}
\usepackage{mathrsfs}
\usepackage{amsfonts}
\usepackage{amsmath,amsfonts,amssymb,mathrsfs,bm}
\usepackage{verbatim,psfrag,times,CJK}
\usepackage{graphicx,graphics,color,epsfig}
\usepackage{float}
\usepackage{flafter}
\usepackage{subeqnarray}
\usepackage{cases}
\usepackage{amsmath, upgreek, amssymb, graphicx, subfigure, paralist, dsfont}
\usepackage[colorlinks, linkcolor=blue, citecolor=blue, urlcolor=blue, breaklinks=true]{hyperref}
\usepackage{subfigure}

\definecolor{RED}{rgb}{1,0,0}
\newcommand{\ket}[1]{\left\vert{#1}\right\rangle}
\newcommand{\bra}[1]{\left\langle{#1}\right\vert}

\begin{document}

\title{Non-classical non-Gaussian state of a mechanical resonator via selectively incoherent damping in three-mode optomechanical systems}

\author{Kang-jing \surname{Huang}$^{1}$}
\author{Yan \surname{Yan}$^{2,3}$}
\author{Jia-pei \surname{Zhu}$^{4,2}$}
\author{Yun-feng \surname{Xiao}$^{1}$}
\author{Gao-xiang \surname{Li}$^{2}$}
\email{gaox@phy.ccnu.edu.cn}

\affiliation{$^{1}$State Key Laboratory for Mesoscopic Physics and School of Physics, Peking University, Beijing 100871, PR China\\
$^{2}$Department of Physics, Huazhong Normal University, Wuhan 430079, PR China\\
$^{3}$Institute of Quantum Optics and Information Photonics, School of Physics and Optoelectronic Engineering, Yangtze University, Jingzhou 434023, PR China\\
$^{4}$Department of Physics, College of Science, Honghe University, 661100 Mengzi, PR China}

\begin{abstract}
We theoretically propose a scheme for the generation of a non-classical single-mode motional state of a mechanical resonator (MR) in the three-mode optomechanical systems, in which two optical modes of the cavities are linearly coupled to each other and one mechanical mode of the MR is optomechanically coupled to the two optical modes with the same coupling strength simultaneously. One cavity is driven by a coherent laser light. By properly tuning the frequency of the weak driving field, we obtain engineered Liouvillian superoperator via engineering the selective interaction Hamiltonian confined to the Fock subspaces. In this case, the motional state of the MR can be prepared into a non-Gaussian state, which possesses the sub-Poisson statistics although its Wigner function is positive.
\end{abstract}


\maketitle
\section{Introduction}
The preparation and quantum control of the nonclassical states of the mechanical resonator (MR) in the cavity optomechanical system has been a subject of longstanding interest\cite{Phystoday58-36} and witnessed a series of developments\cite{annphys525-215, jpb57-36, revmodphys86-1391}. They have potential prospect in the ultraprecision measurement\cite{physrep511-273}, long-distance quantum communication and networking\cite{nature453-1023, prl92-013602, prl92-197901}, and so on. Furthermore, nonclassical states of the MR have been already investigated in various optomechanical systems. For instance, achieving squeezed states in the MR can be realized by putting an optical parametric amplifier inside a cavity\cite{pra79-013821}, or by quantum state transferring from squeezed light driving the cavity via dispersive coupling\cite{pra79-063819}, or via the destructive interference of quantum noise via dissipative coupling\cite{pra88-013835}. And a recent experiment has investigated that a stationary quadrature-squeezed state of the MR is produced by using microwave frequency radiation pressure\cite{science349-952}. A theoretical proposal for preparing the entangled mechanical ¡®cat¡¯ state via conditional single-photon optomechanics has been presented\cite{njp15-093007}. The stationary entanglement between the mechanical modes can be prepared by using optimized two-tone (or four-tone) driving of a cavity only with one driven auxiliary mode as the engineered reservoir\cite{pra89-063805}, or hybird quantum interfaces\cite{prl109-223601}, or suitable intensity modulation of a single laser beam in the optomechanical systems\cite{njp14-125005}, or sideband exicitations\cite{pra88-022325} in the optomechanical systems.

However, most of the theoretical schemes for preparing the non-classical states of the MRs in the cavity optomechanical systems focus on the generation of Gaussian states. It is well known that Gaussian states play a prominent role in the quantum information processing with continuous variables because they are easy to handle via being fully described by the first and second moments of their covariance matrix and their evolution under quadratic Hamiltonians is easily cast in the phase space\cite{prl84-2726}. Indeed, non-Gaussian states might be more appropriate than Gaussian state in view of the robustness when considering long distance communication. Multi-photon squeezed states are a kind of typical continuous variable non-Gaussian states, which can be generated by considering nonlinear extensions of the linear Bogoliubov squeezing transformations\cite{pra69-033812}. And quantum superposed states are another kind of typical continuous variable non-Gaussian states. In particular, macroscopical quantum superposed states of the MR have been studied in both experiments and theories. The NOON state of the MR with arbitrary phonon numbers is generated via manipulating photon transport in the single-photon strong-coupling optomechanical systems\cite{pra87-033807}. Under a resonance condition in the resolved-sideband limit, the coherent superposition states of the MR is prepared when the optomechanical cavity is decoupled from the external field due to quantum interference\cite{pra87-053849}. Similarly, via utilizing the quantum interference, the MR can evolve into a squeezed non-Gaussian state beyond the resolved-sideband limit, where Wigner function of the motional of the MR presents negative values\cite{oe22-018254}. In addition, the quantum superposition between the vacuum and an arbitrary Fock state of the MRs is created via strong single-photon optomechanical coupling, which also results from quantum interference\cite{pra89-053829}.Recently, Xin-you L\"{u} et al. utilize the nonlinear interaction between a squeezed cavity mode and a mechanical mode in an optomechanical system to engineer nonclassical phonon states\cite{prl114-093602}.

Recent experiments show that the MRs can be cooled to their quantum ground states\cite{nature444-67, nature444-71, nature478-89}. In the strong optomechanical coupling regime, the cooling process can be significantly accelerated by controling the cavity dissipation\cite{110-153606}. And the optomechanical systems are appraoching the strong single-photon coupling regime\cite{nature452-72, naturephys6-707, nature475-359}. Thus the optomechanical systems provide a very good platform to explore the macroscopic quantum states of the MRs. And the quantum superpositions are main resources for quantum information processing\cite{oe21-5529}. Motivated by these reseaches, we use a three-mode optomechanical system to study the nonclassical phonon states engineering.

In this paper, we propose a new scheme to prepare the non-Gaussian state of the MR when a coherent laser field is applied to a three-mode cavity optomechanical system, in which two optical modes of the cavities are linearly coupled to each other and one mechanical mode of the MR is optomechanically coupled to the two optical modes simultaneously. We confine the interaction Hamiltonian in the Fock subspaces and obtain the engineered Liouvillian superoperator. By using the incoherent pumping of the phonon reservoir, we can get the motional state of the MR prepared in the non-classical non-Gaussian state. The steady state of the MR posses the non-Gaussianity and the non-classicality as its non-Gaussianity is greater than zero and its SOCF is less than 1.

The paper is organized as follows. In Sec.~\ref{sec2}, the model is introduced and the effective master equation of the MR is derived. In Sec.~\ref{sec3} we study the properties of the stationary motional state of the MR. In Sec.~\ref{sec4}, we summarize our results.

\section{Description of the system}\label{sec2}
We consider a three-mode optomechanical system where a single-mode mechanical resonator (MR) with frequency $\omega_m$ is simultaneously coupled to two single-mode cavities with resonance frequencies $\omega_a$ and $\omega_b$ via dispersive couplings, as shown in Fig. 1.
\begin{figure}[hpt]
\centering\includegraphics[width=6cm,keepaspectratio,clip]{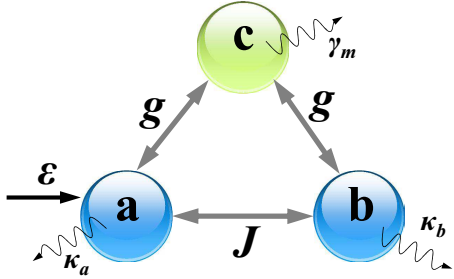}
\caption{(Color Online) The schematic of the three-mode optomechanical system consisting of two optical modes ($a$ and $b$) and one mechanical mode ($c$). The cavity $b$ is linearly coupled to the cavity $a$ which is driven by an external weak laser field. And the two cavities couple to the MR via radiation pressure with the same coupling strength $g$, respectively.}
\label{fig1}
\end{figure}
The cavity $b$ couples to the cavity $a$ with the coupling strength $J$, which is driven by an external coherent light fields with frequency $\omega_{L}$ at a rate $\varepsilon$\cite{pra91-033818, pra91-053854}. In the frame rotating at the input laser frequency $\omega_{L}$, the total Hamiltonian of this system can be expressed in the form after performing a rotating-wave approximation on the driving term (setting $\bar{h}=1$ throughout this paper)
\begin{align}
H=H_0+H_1\label{H},
\end{align}
where $H_0$ is the free Hamiltonian of the total system, which contains the two cavities and the MR, and reads
\begin{align}
H_0=\Delta_aa^{\dag}a+\Delta_bb^{\dag}b+\omega_mc^{\dag}c.
\end{align}
Here, $a$, $b$, and $c$ are the annihilation operators of the two cavity modes and the MR mode. $\Delta_j=\omega_j-\omega_L(j=a,b)$ are the detunings of the laser field from the resonant frequency of the cavity $j$. The interaction Hamiltonian $H_1$ in Eq.~\eqref{H} is given by
\begin{align}
H_1\!=\!g(a^{\dag}a\!+\!b^{\dag}b)(c\!+\!c^{\dag})\!+\!J(a^{\dag}b\!+\!ab^{\dag})\!+\!\varepsilon (a\!+\!a^{\dag}),\label{H1}
\end{align}
where the parameters $g$ is the single-photon optomechanical coupling strength and we have assumed the two optomechanical coupling strengths are the same.

In order to get rid of the terms involving the cavity fields' intensities and the MR, a polaron transformation is needed\cite{prl107-063601} and it displaces the MR. The transformed Hamiltonian is defined by
\begin{align}
\tilde{H}=e^{S}He^{-S},
\end{align}
where $S=\eta (a^{\dag}a+b^{\dag}b)(c^{\dag}-c)$, and we are setting $\eta=g/\omega_m$. The Lamb-Dicke parameter $\eta$ measures the localization of the motional ground state of the MR relative to the (effective) wavelength of the transition. After this transformation, the total Hamiltonian of the system is of form
\begin{eqnarray}
\tilde{H}=H_0+\tilde{H}_1,
\end{eqnarray}
where
\begin{eqnarray}
\tilde{H}_1=J(a^{\dag}b\!+\!ab^{\dag})+\varepsilon(a^{\dag}e^{\eta (c^{\dag}-c)}\!+\!\textrm{H.c.})\label{wwH},
\end{eqnarray}
describes the Hamiltonian of a modified sysytem (the MR plus the two cavities). The first term of the Hamiltonian $\tilde{H}_1$ in Eq.~\eqref{wwH} remain the same after the polaron transformation, the second term represents the interaction among the MR, the cavity $a$ and the driving field. During the process, we have ignored the effective Kerr-type nonlinearity. In the absence of driving terms, the photon and the phonon degrees of freedom in the optomechanical system can be decoupled by means of a polaron transformation\cite{prl107-063601}.

Since the two non-commuting operators $A=e^{\eta c}$ and $B=e^{\eta c^{\dag}}$ satisfy the condition of $[A, [A, B]]=[B, [A, B]]=0$, the Baker-Campbell-Hausdorff Relation $e^{\alpha (A+B)}=e^{\alpha A}e^{\alpha B}e^{-\frac{\alpha^2}{2}[A, B]}$ can be applied. And we expanse it by using a Taylor's series method and obtain
\begin{eqnarray}
e^{\alpha (A+B)}=e^{\frac{\eta^2}{2}}\sum_{m,n=0}^{\infty}\frac{\eta^m(-\eta)^n}{m!n!}c^nc^{\dag^m},
\end{eqnarray}
Again, we neglect the processes involving two- and multi-phonon absorption or emission such that the Taylor series expansion can be applied to approximately rewrite, i.e., we only keep the terms $n=m$, $n=m+1$, and $n=m-1$. So the Hamiltonian of the whole system possesses the terms $c^mc^{\dag^m}$,  $c^{m+1}c^{\dag^m}$, and $c^mc^{\dag^{m+1}}$. In this case, the interaction Hamiltonian $\tilde{H}_1$ is transformed to the interaction picture with the unitary transformation $U=e^{iH_0t}$, and find
\begin{align}
V\!=&e^{\frac{\eta^2}{2}}\varepsilon a^{\dag}e^{i\Delta_a t}[f_1(cc^{\dag})\!-\!cf_2(cc^{\dag})e^{-i\omega_mt}\!+\!f_2(cc^{\dag})c^{\dag}e^{i\omega_mt}]\nonumber\\
&+Ja^{\dag}be^{i(\Delta_a-\Delta_b)t}+\textrm{H.c.}
\end{align}
where $f_1(cc^{\dag})=\sum_{m=0}^{\infty}\frac{(-1)^m\eta^{2m}}{(m!)^2}c^mc^{\dag^m}$, $f_2(cc^{\dag})=\sum_{m=0}^{\infty}\frac{(-1)^m\eta^{2m+1}}{m!(m+1)!}c^mc^{\dag^m}$.

In this paper, we are interested in the case that there is at most one photon in the two cavities. So the annihilation operators $a$ and $b$ of the two cavities can be expanded in terms of Fock states as $\hat{A}=\sum_{n,n'=0}^1\ket{n}\bra{n}\hat{A}\ket{n'}\bra{n'}=\ket{0}\bra{1}$ ($\hat{A}=a,b$). And $\ket{m_a, n_b}$ represents the Fock state with $m$ particles in the cavity $a$ and $n$ particles in the cavity $b$. With the weak driving field, there is no more than one particle in the two cavities. we can adopt the new operators to express the truncated Fock space as $\ket{e}=\ket{0_a, 1_b}$, $\ket{i}=\ket{1_a, 0_b}$ and $\ket{g}=\ket{0_a, 0_b}$. Obviously, the states $\ket{e}$ and $\ket{i}$ are the single photon state, the state $\ket{g}$ is the ground state of the two cavities. By using the new eigenbasis of the Fock states, the interaction Hamiltonian $V$ becomes
\begin{align}
V\!=&J\ket{i}\bra{e}e^{i(\Delta_a-\Delta_b)t}+e^{\frac{\eta^2}{2}}\varepsilon\ket{i}\bra{g}e^{i\Delta_a t}[f_1(cc^{\dag})\!\nonumber\\
&-\!cf_2(cc^{\dag})e^{-i\omega_mt}\!+\!f_2(cc^{\dag})c^{\dag}e^{i\omega_mt}]+\textrm{H.c.}\label{V}
\end{align}

Next, we assume that the MR and the driving laser are significantly detuned from the transition frequencies of the new three states: $J\ll\Delta_a-\Delta_b$, $\varepsilon\ll\Delta_a$, and $\eta\varepsilon\ll\Delta_a\pm\omega_m$. Under this large detuning condition, we choose that the detuning $\Delta_b$ between the laser field frequency and the resonant frequency of the cavity $b$ satisfies this condition $\Delta_b=\omega_m$ and we can perform the standard adiabatic eliminate\cite{pra73-043803, cjphys85-625} of the single photon state $\ket{i}$ and obtain an effective Hamiltonian $H_{\textrm{eff}}=-iV(t)\int V(t')\textrm{d}t'$, which describes the dynamics of the system containing the MR and the two states $\ket{g}$, $\ket{e}$. During this calculation, we apply the rotating wave approximation to discard the high oscillatory terms and obtain the second-order effective Hamiltonian
\begin{align}
H_{\textrm{eff}}\!&=\!H'_0+\!(\alpha c^{\dag}\ket{g}\bra{e}\!+\!\textrm{H.c.})\label{Heff},
\end{align}
where the parameter is $\alpha\!=\!\frac{e^{\frac{\eta^2}{2}}J\varepsilon}{\Delta_a-\omega_m}$, and
\begin{align}
H'_0=&\chi_e\ket{e}\bra{e}\!-\!e^{\eta^2}\varepsilon^2[\frac{1}{\Delta_a}f_1^2(cc^{\dag})\!+\!\frac{1}{\Delta_a\!+\!\omega_m}cf_2^2(cc^{\dag})\nonumber\\
&+\!\frac{1}{\Delta_a\!-\!\omega_m}f_2(cc^{\dag})c^{\dag}cf_2(cc^{\dag})]\ket{g}\bra{g}\label{H00},
\end{align}
corresponds to dynamical energy shifts of $\ket{e}$ and $\ket{g}$ due to the action of the driving field.
The parameter $\chi_e$ in Eq.~\eqref{H00} is $\chi_e=\frac{J^2}{\omega_m-\Delta_a}$. Note that the energy shift of $\ket{g}$ depends obviously on the number of phonons in the MR. And the difference of energy shifts of $\ket{g}$ and $\ket{e}$ will determine the effective resonance frequency of the $\ket{e}\leftrightarrow\ket{g}$ transition. $\alpha$ and $e^{\eta^2}\varepsilon^2$ represent on- and off-resonant couplings between the MR and the $\ket{e}\leftrightarrow\ket{g}$ transition. To get selectivity, we adopt a reference frame defined by the unitary transformation $U=e^{-iH'_0t}$ and the relation $c^{\dag}=\sum_{n=1}^{\infty}\sqrt{n+1}\ket{n+1}\bra{n}$ to deduce the desired selective interaction. Thus, we can get rid of the diagonal terms in Eq.~\eqref{Heff} and focus on the non-diagonal interaction terms with the simplified form
\begin{align}
V_{\textrm{eff}}\!=\!\alpha_n\ket{g,n+1}\bra{e,n}e^{i\varphi_nt}\!+\!\textrm{H.c.},\label{Veff}
\end{align}
where $\alpha_n\!=\!\sum\limits_{\substack{m\!=\!0\\n\!=\!1}}^{\infty}\eta\sqrt{n+1}g(m,n)$ and the time-dependent phase factor
\begin{align}
\varphi_n\!=&\!-\!\chi_e-e^{\eta^2}\varepsilon^2\sum_{\substack{m,k\!=\!0\\n=1}}^{\infty}\{g(m,n)g(n,k)[\frac{1}{\Delta_a}\prod_{x=m,n,k}(x+1)\nonumber\\
&\!+\!\frac{\eta^2}{\Delta_a\!-\!\omega_m}\!+\!\frac{\eta^2\prod\limits_{x=m,k}(n+x+2)}{(\Delta_a\!+\!\omega_m)\prod\limits_{x=1,2}(n+x)}]\}
\end{align}
with the function $g(x,y)=\frac{(-1)^x\eta^{2x}(x+y+1)!}{x!(x+1)!(y+1)!}$. The Hamiltonian $V_{\textrm{eff}}$ is block separable in the subspaces spanned by the states ${\ket{g,n},\ket{e,n+1}}$ of the system. So far, we have obtained the Jaynes-Cummings Hamiltonian. In order to restrict the interaction Hamiltonian to the subspace ${\ket{j}, \ket{j+1}}$, we impose the simultaneous conditions $\varphi_j=0$ and $\varphi_n\gg|\alpha_n| (n\neq j)$. Under the restriction, we can eliminate all the transitions to adjacent levels ${\ket{n}, \ket{n+1}}$ except when $n=j$ with the rotating wave approximation and obtain the sliced Hamiltonian
\begin{align}
V_{k}\!=\!{\alpha_j\ket{g, j+1}\bra{e, j}\!+\!\textrm{H.c.}}\label{Vk},
\end{align}
It is worth noting that the effective Rabi frequency $\alpha_j$ increases with the selected excitation $j$ of the MR mode. From the obtained interactive Hamiltonian $V_{j}$ in Eq.~\eqref{Vk}, we can see that the desired selective transition $g\leftrightarrow e$ is restricted in the subspace $\{\ket{j}, \ket{j+1}\}$.

We assume that the interactions between the two cavities and the MR are weak and can trace over the operators of the cavities in the sliced Hamiltonian $V_{j}$ in Eq.~\eqref{Vk} and calculate the master equation of the reduced density matrix of the MR within the Born-Markov approximation. In addition, the coupling constant $\alpha_j$ is smaller than the characteristic rates determining the dynamics of the two cavities. So we can know that both the cavities are in the vacuum state $\ket{0_a, 0_b}$, i.e., $\ket{g}$ when they arrive the steady state. In this regime, the photon degrees of the cavities are in the steady state. Then, when we consider the influence of the phonon reservoir, the reduced density operator $\rho$ of the MR, using the Born-Markov approximations, satisfies the following equation
\begin{align}
\frac{\textrm{d}\rho}{\textrm{d}t}=\frac{\alpha_j^2}{\kappa_b}\mathcal{D}(c_j)+\frac{\gamma_p}{2}(\bar{n}_p+1)\mathcal{D}(c)+\frac{\gamma_p}{2}\bar{n}_p\mathcal{D}(c^{\dag})\label{rho},
\end{align}
where $\mathcal{D}(\mathcal{O})\!=\!2\mathcal{O}\rho \mathcal{O}^{\dag}\!-\!\rho \mathcal{O}^{\dag}\mathcal{O}\!-\!\mathcal{O}^{\dag}\mathcal{O}\rho$ and $\kappa_b$ is the decay of the cavity $b$. $c_j\!=\!\ket{j}\bra{j+1}$ represents a selective annihilation operator and the first term of Eq.~\eqref{rho} stands for the emission. The second and last terms in Eq.~\eqref{rho} stand for the processes leading to thermalization at rate $\gamma_p$ with a thermal environment at temperature $T_p$, where the mean number of phonon at frequency $\omega_p$ equals to $\bar{n}_p=[\text{exp}(\frac{\omega_p}{k_BT_p})-1]^{-1}$.

\section{Results and discussions}\label{sec3}
Until now, we have obtained the effective master equation which is able to describe the dynamics of the phonon. In this section, we shall investigate the properties of the stationary state of the phonon.

The phonon Fock state occupation probabilities are given by
\begin{equation}\label{rhonn}
\rho_{nn}=\begin{cases}
\zeta_n\rho_{00}, &(n\leq j);\\
\zeta_n\varpi_j\rho_{00}, &(n>j).
\end{cases}
\end{equation}
which can be obtained from Eq.~\eqref{rho}. Here the parameters $\zeta_n$, $\varpi_j$, and $\rho_{00}$ are given by
\begin{align}
\zeta_n&=(\frac{\bar{n}_p}{\bar{n}_p+1})^n,\\
\varpi_j&=\frac{\gamma_p(j+1)}{\gamma_p(j+1)+\epsilon_j},\\
\rho_{00}&=\frac{1}{(\bar{n}_p+1)(1-\zeta_{j+1}+\varpi_j\zeta_{j+1})}.
\end{align}
with $\epsilon_j=\frac{2\alpha_j^2}{\kappa_b(\bar{n}_p+1)}$. It's worth noting that the Fock state space of the MR can be restricted up to $\ket{j}$ through the appropriate choice of the parameter $\varpi_j$, which may be very small so that the condition $\zeta_{j+1}\varpi_j\ll\zeta_j$ is satisfied. Typically, for a three-mode optomechanical system based on whispering-gallery cavities, we can choose $\omega_m=10g$, $J=g$, $\varepsilon=3g$, $\bar{n}_p=10$, $\gamma_p=10^{-5}g$, $\kappa_b=0.15g$, $\Delta_b=\omega_m$.

\subsection{Preparation of steady Fock states of the MR}\label{sec3b}
\begin{figure}
  \centering
  \subfigure[$\eta=0.1$]{
    \label{fig2a} 
    \includegraphics[width=5cm]{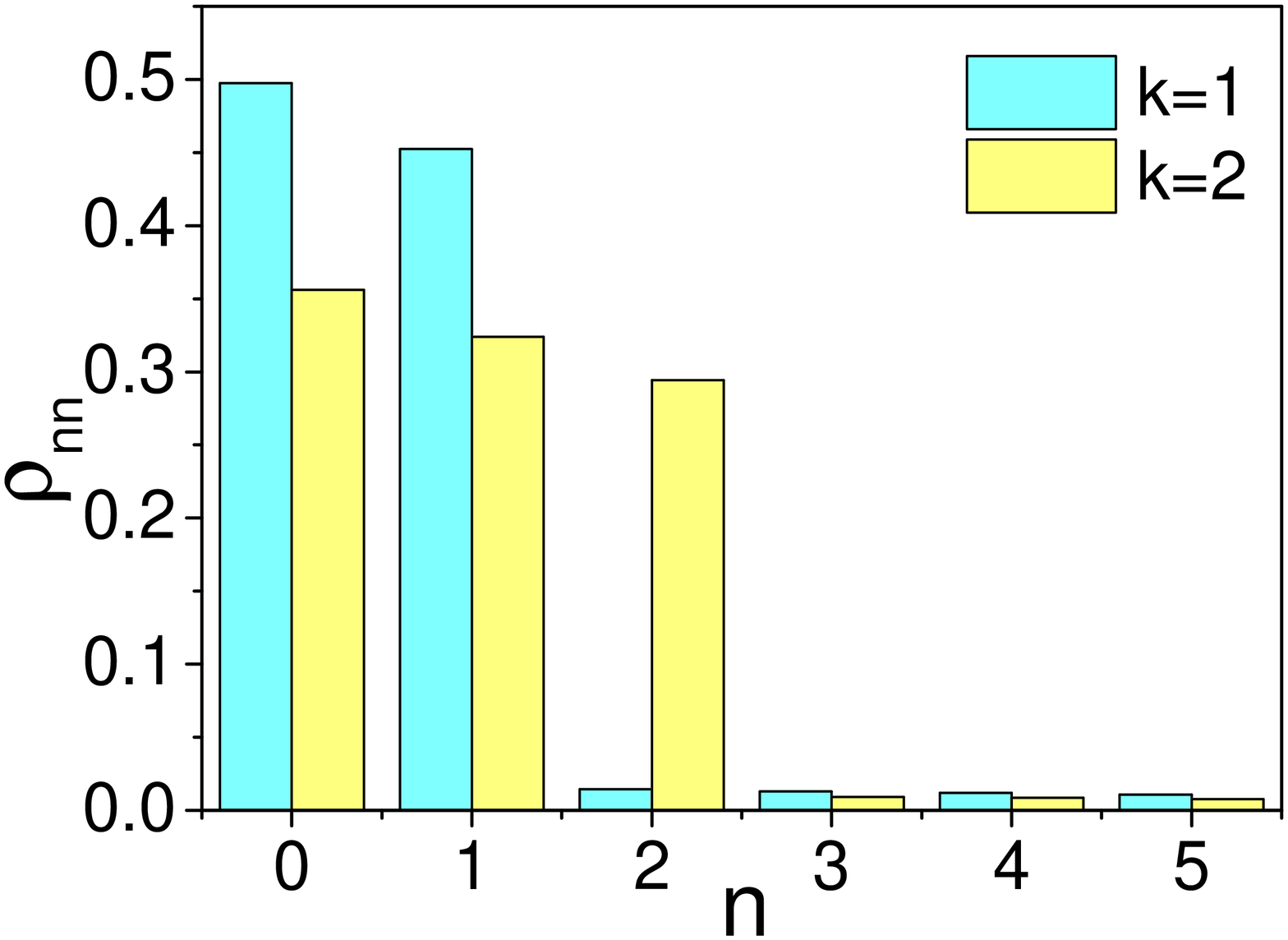}}
  \hspace{0.5cm}
  \subfigure[$\eta=0.3$]{
    \label{fig2b} 
    \includegraphics[width=5cm]{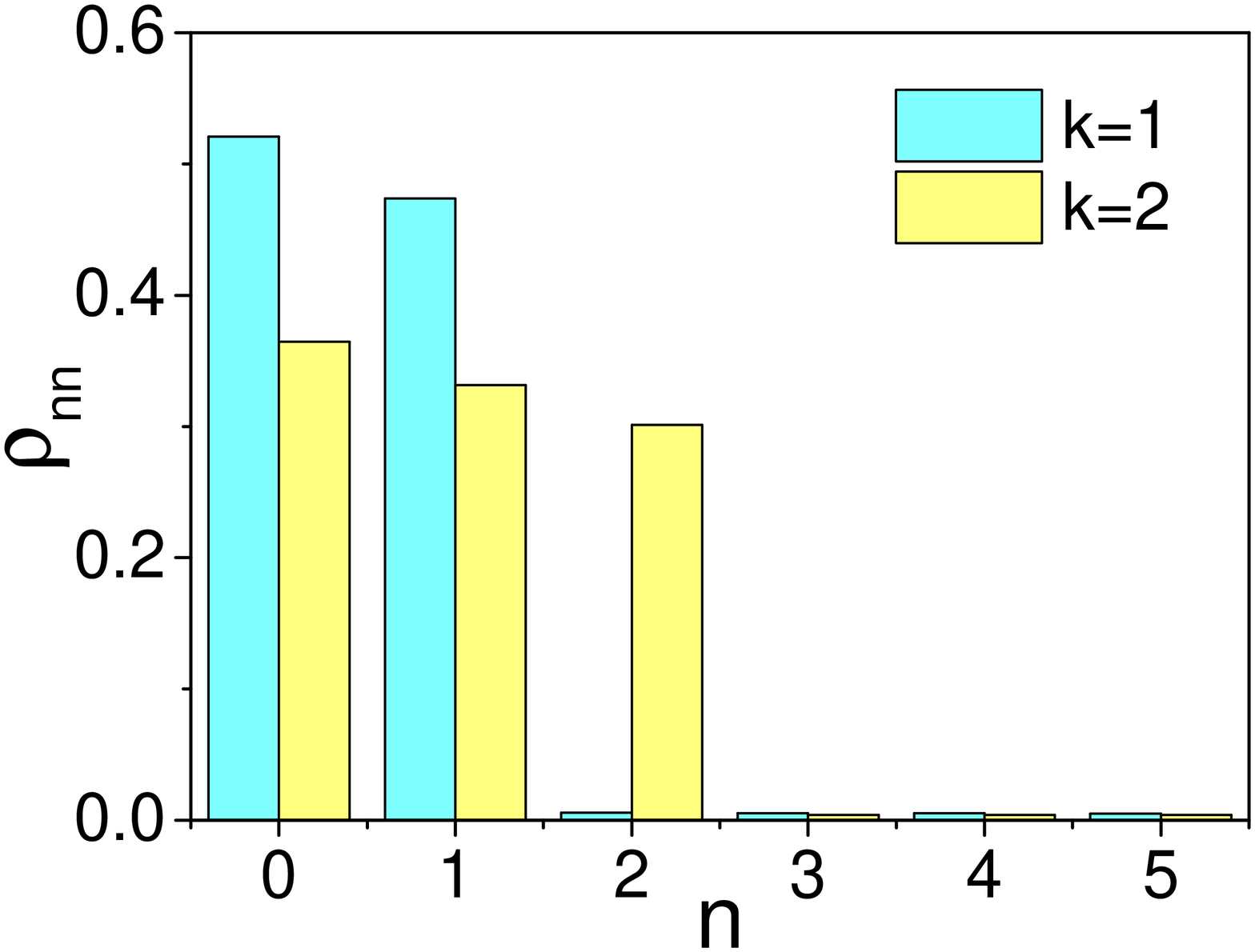}}
  \caption{The phonon number distribution of the MR in the steady state regime versus $n$ for different values of $\eta$ and $k$. (a) the cyan corresponds to $j=1$ (in this case, $\Delta_a=-9.7g$); the yellow corresponds to $j=2$ (in this case, $\Delta_a=-9.6g$). (b) the cyan corresponds to $j=1$ (in this case, $\Delta_a=-7.5g$); the yellow corresponds to $j=2$ (in this case, $\Delta_a=-6.6g$).The other parameters are as follows: $\omega_m=10g$, $J=g$, $\varepsilon=3g$, $\gamma_p=10^{-5}g$, $\kappa_b=0.15g$, $\Delta_b=10g$, $\bar{n}_p=10$.}
  \label{fig2} 
\end{figure}

In order to generate the steady MR state, a significantly large $\epsilon_k$ is needed. In this case, $\varpi_j$ is small enough compared to the coefficient $(j+1)\gamma$, so that $\zeta_{j+1}\varpi_j\ll\zeta_j$ entails the truncation of the number state population $\rho_{nn}$ from the population of state $j+1$. With the two different Lamb-Dicke parameter values $\eta=0.1$ and $\eta=0.3$, we choose the two truncated Fock state space $j=1$ or $j=2$ and other parameters in accordance with the condition of the truncation of number state population. In Fig.~\ref{fig2}, we show the phonon number distribution of the MR with different parameters in the steady state. It shows that the probabilities of $n$ ($n\leq j$ ) phonon detection are nearly equal, but the probability of generating more than one (or two) phonon pair is very small and nearly zero when $j=1$ (or 2). When the Fock state space is given, the parameter $\varpi_j$ increases and the phonon Fock state occupation probabilities $\rho_{nn}(n>j)$ decrease along with the increasing of $\eta$. At the same time, $\rho_{nn}(n<j)$ increase. So we can clearly see from Fig.~\ref{fig2} that $\rho_{00}$ in Fig.~\ref{fig2a} is slightly smaller than in Fig.~\ref{fig2b}.

As a matter of fact, the mixed state of the MR which can be prepared results from incoherent population pumping. In the absence of the incoherent pump, so $\alpha_n=0$, the steady state $\rho_{nn}$ in Eq.~\eqref{rhonn} equals the thermal state. The population is then thermally distributed among the whole Fock state space. After switching on the incoherent pump, it is easily to find from Eq.~\eqref{rho} that the first term indicates the processes which can decrease one excitation of the MR in the Fock state space $\{\ket{j}, \ket{j+1}\}$ and the second and third terms establish the balance of the system. Then, the steady MR state is almost confined to the Fock state space $\{\ket{0}, \ket{1},\cdots, \ket{j}\}$. In this case, there is nearly zero populations in the Fock state $\ket{l} (l>j)$. With the help of incoherent population pumping, the MR is trapped in the mixed state in the low Fock basis.

\subsection{Wigner function of the MR and non-Gaussianity}\label{sec3b}

The Wigner function (WF) of a state $\rho$ is a quasi-probability distribution which fully describes the states of a quantum system in phase space. It is a useful tool to study the non-classical properties of quantum states, because of the partial negativity of the WF~\cite{wiger, kenfack} is a good indication of the non-classical character of the state. According to Hudson¡¯s theorem \cite{mathphys6-249}, all pure states with negative Wigner function are non-Gaussian. But the situation is more complex for mixed states. In the following, we shall exhibit the WF of the MR and discuss whether the mixed state of the MR is non-Gaussian.

\begin{figure}
\centering
\subfigure[$k$=1]{
\label{fig4} 
\includegraphics[width=7cm]{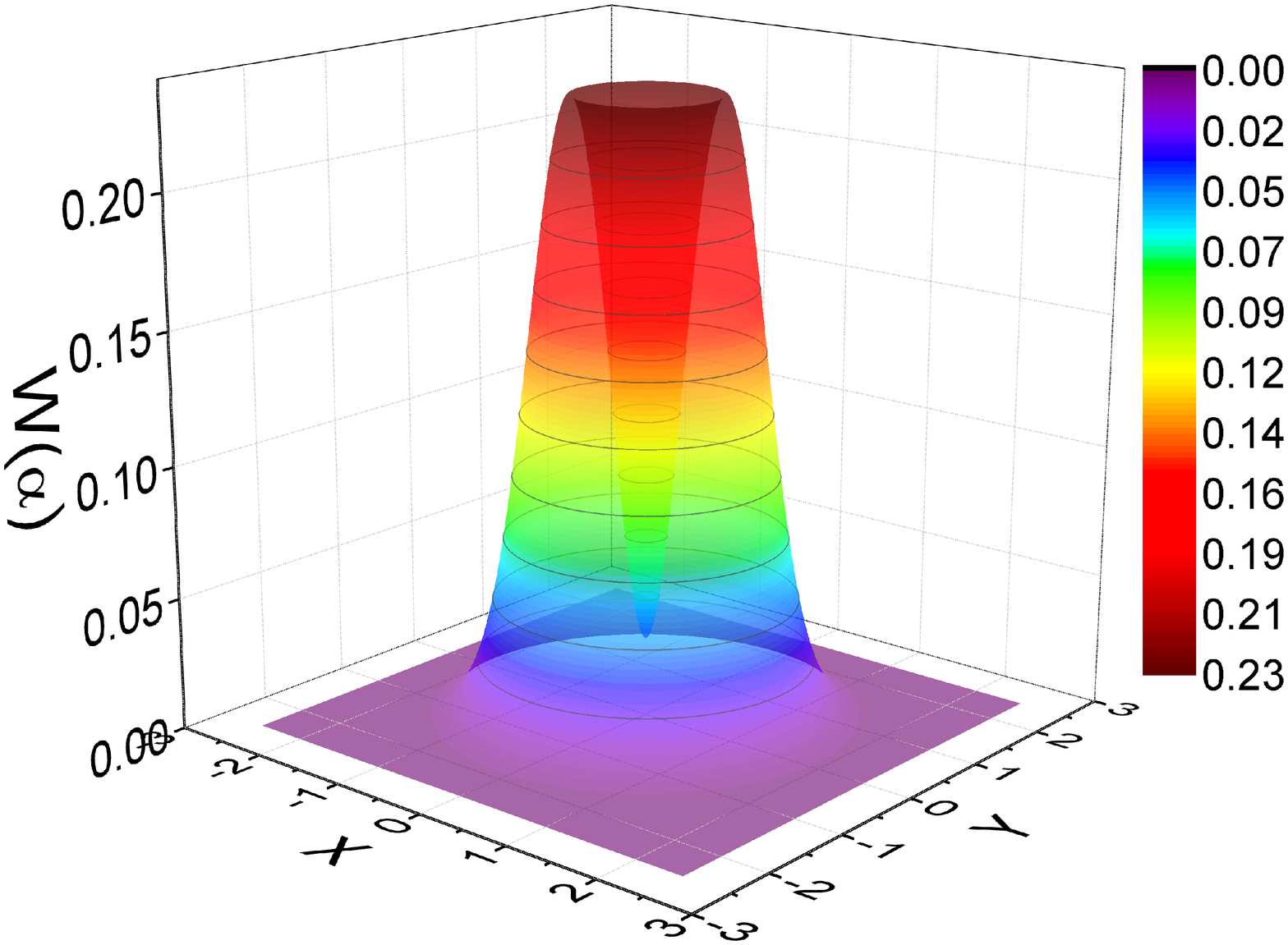}}
\hspace{1in}
\subfigure[$k$=2]{
\label{fig5} 
\includegraphics[width=7cm]{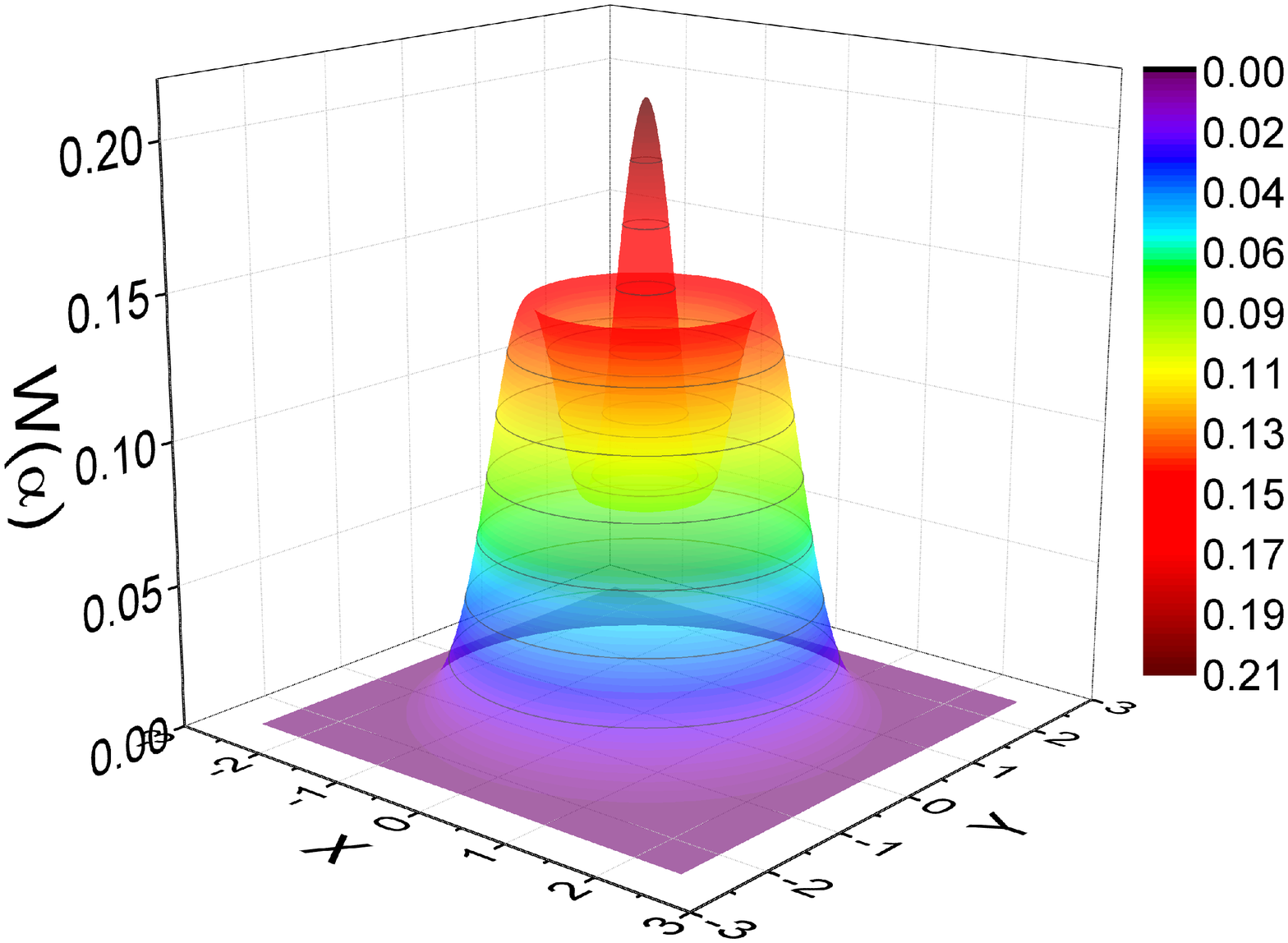}}
\caption{The WF of the MR in the steady state regime versus $n$ for different value of $k$. (a) $k=1$. (b) $k=2$. Other parameters are the same with that in Fig.~\ref{fig2b}.}
\label{fig3} 
\end{figure}

From Eq.~\eqref{rho}, we can know that there is not coherent term but only incoherent terms in the system, i.e., $H=0$. The off-diagonal elements of the MR's reduced density matrix reflect on the quantum coherence properties of the system, so there are only diagonal elements in the reduced density matrix of the system evolution. Then, when the state of the MR is expressed as the superposition of the number state $\rho=\sum_{n=0}^{\infty}\rho_{nn}$, the WF of the state $\rho$ of the MR is found to be
\begin{align}
W(\xi)=\frac{2}{\pi}\sum_{n=0}^{\infty}(-1)^ne^{-2|\xi|^2}\rho_{nn}L_n(4|\xi|^2),\label{W}
\end{align}
where $L_n(x)$ is the Laguerre polynomial of order $n$. As is shown in Eq.~\eqref{W}, we can see clearly that, to obtain the WF of the phonon on each point of the phase space, all one needs to know is the number distribution of the MR, after it has been displaced in the phase space. We can use the selective scheme discussed above to measure $\rho_{nn}$ and then, Eq.~\eqref{W} to calculate $W(\xi)$. The WF of the truncated distribution is plotted in Fig.~\ref{fig3} in phase space when $\eta=0.3$, where $\bar{X}$ and $\bar{Y}$ are the position and the momentum.

The phonon which is always in a mixed state when it arrives at the steady state in our system is described by the entirely positive WF. Therefore, we are unable to estimate the non-classical properties of the phonon in the WF of Fig.~\ref{fig3}. Whether the phonon possesses the non-classical property will be discussed in other ways later. By looking into the Fig.~\ref{fig3}, we can see that the Wigner distribution does not exhibit bell shape, so we can know that the mixed state of the MR is non-Gaussian. But how to quantify the non-Gaussian character of a quantum state? Based on the Hilbert-Schmidt distance between $\rho$ and a reference Gaussian state $\rho_G$, we introduce the non-Gaussianity $\delta[\rho]$ of a quantum state\cite{pra76-042327, physscriptaT153-014028}
\begin{equation}
\delta[\rho]=\frac{1}{2}[1+\frac{\textrm{Tr}(\rho_G^2)-2\textrm{Tr}(\rho_G\rho)}{\textrm{Tr}(\rho^2)}],
\end{equation}
which can quantify how much a state fails to be Gaussian. The Gaussian state $\rho_G$ is the same covariance matrix and the same vector of the state $\rho$. Specially, for any mixed Fock-diagonal state, the Gaussian reference state $\rho_G$ is a thermal state with the same mean occupancy of the mixed Fock-diagonal state. For the mixed Fock-diagonal state, the non-Gaussianity $\delta[\rho]$ is given by an easily computable expression
\begin{equation}
\delta[\rho]=\sum_{n=0}^{\infty}\rho_{nn}\textrm{ln}\rho_{nn}+(\bar{n}+1)\textrm{ln}(\bar{n}+1)-\bar{n}\textrm{ln}\bar{n},
\end{equation}
where $\bar{n}$ is the mean phonon number of the MR. By simply calculating, we can obtain the results $\delta[\rho]=0.15 (j=1, \eta=0.1)$, $0.18 (j=2, \eta=0.1)$, $0.22 (j=1, \eta=0.3)$, $0.23 (j=2, \eta=0.3)$. According to the lemmas in Ref.\cite{pra76-042327, pra78-060303}, we can know that $\delta[\rho]$ is a well-defined non-negative quantity and equals zero if and only if $\rho$ is a Gaussian state. And all of our results are greater than zero, so we can say that the mixed state of the MR is a non-Gaussian state.

\subsection{Nonclassical two-order phonon correlations}\label{sec3c}
From the discussion above, we have known that the mixed state of the MR is non-Gaussian and its WF is positive. We can not observe the non-classical property of the MR. In the following, we will discuss the non-classical property of the mixed state of the MR from the point of the second-order correlation function (SOCF) $g^2(\tau)=\frac{\langle b^{\dag}b^{\dag}(\tau)b(\tau)b\rangle}{\langle b^{\dag}b\rangle^2}$ at $\tau=0$\cite{g20}, where $b^{\dag}(b)$ is the creation (annihilation) operator of the phonon. Quantum correlations surely play a role in one way or the other in quantum computers and simulators outperforming their classical counterpart. This SOCF quantifies, how the detection of one phonon from a phonon source influences the probability to detect another one. The SOCF $g^{(2)}(0)$ already allows one to distinguish between super-Possonian ($g^{(2)}(0)>1$), Possonian($g^{(2)}(0)=1$) and sub-Possonian($g^{(2)}(0)<1$) phonon statistics distribution. The property $g^{(2)}(0)<1$ is characteristic of nonclassical state. For attaining more information about the phonon number distribution and further discussing the nonclassical properties of the phonons, we calculate the SOCF of the phonons. The SOCF is
\begin{align}
g^{(2)}{(0)}=\frac{\sum_{n=0}^{\infty}{n(n-1)\rho_{nn}}}{(\sum_{n=0}^{\infty}{n\rho_{nn}})^2}.\label{g2}
\end{align}
If $g^{(2)}(0)<1$ is satisfied, the phonon counting distribution is narrower than a Poissonian one, which implies that the phonon is referred to as sub-Poissonian. With the Lamb-Dicke parameter value $\eta=0.1$, we can calculate the SOCF $g^{(2)}{(0)}=0.51$ ( or $0.44$) when the truncated Fock state space is $j=1$ ( or $2$). For another Lamb-Dicke parameter value $\eta=0.3$, the value of the SOCF is $g^{(2)}{(0)}=0.65$ ( or $0.48$) when $j=1$ ( or $2$). In our system, the SOCFs are less than 1. The fact $g^{(2)}{(0)}<1$ indicates the sub-Poissonian statistics and the non-classicality of the steady state of the MR. It also implies that the generated state of the MR exhibits the sub-Poissonian behaviour and non-classical effect.

\section{Conclusions}\label{sec4}
To summarize, we have proposed a scheme to prepare a non-classical single-mode motional state of the MR in the three-mode optomechanical systems where two optical modes of the cavities are linearly coupled to each other and one mechanical mode of the MR is optomechanically coupled to the two optical modes simultaneously. Our proposal relies on engineering the selective interaction Hamiltonian confined to the Fock subspaces, which can generate the engineered Liouvillian superoperator to govern the MR dynamics. Then, the motional state of the MR is prepared in the steady non-classical non-Gaussian state as its non-Gaussianity is greater than zero and it exhibits the non-classicality when its SOCF shows the sub-Poissonian distribution although its WF approaches the positive region.

\section*{Acknowledgements}
This work is supported by the National Natural Science Foundation of China (Grant No. 61275123 and No.11474119) and the National Basic Research Program of China (Grant No. 2012CB921602).

\end{document}